\documentstyle[psfig]{aipproc}

\def\be{\begin{equation}}
\def\ee{\end{equation}}
\def\lsim{\lower 2pt \hbox{$\, \buildrel {\scriptstyle <}\over
         {\scriptstyle \sim}\,$}}
\def\gsim{\lower 2pt \hbox{$\, \buildrel {\scriptstyle >}\over
         {\scriptstyle \sim}\,$}}

\begin{document}
\title{Gamma-Ray Pulsars: \\Models and Predictions}

\author{Alice K. Harding}
\address{NASA Goddard Space Flight Center, Greenbelt MD 20771, USA}

\maketitle

\begin{abstract}
Pulsed emission from $\gamma$-ray pulsars originates inside the magnetosphere, 
from radiation by 
charged particles accelerated near the magnetic poles or in the 
outer gaps.  In polar cap models, the high energy spectrum is cut off 
by magnetic pair production above an energy that is dependent on the 
local magnetic field strength.  While most young pulsars with surface 
fields in the range $B = 10^{12} - 10^{13}$ G are expected to have high energy 
cutoffs around several GeV, the gamma-ray spectra of old pulsars having 
lower surface fields may extend to 50 GeV.  Although the gamma-ray 
emission of older pulsars is weaker, detecting pulsed emission at high 
energies from nearby sources would be an important confirmation of polar 
cap models.  Outer gap models predict more gradual high-energy turnovers 
at around 10 GeV, but also predict an 
inverse Compton component extending to TeV energies.  Detection of 
pulsed TeV emission, which would not survive attenuation at the polar 
caps, is thus an important test of outer gap models.  Next-generation 
gamma-ray telescopes sensitive to GeV-TeV emission will provide critical 
tests of pulsar acceleration and emission mechanisms.

\end{abstract}

\section*{Introduction}

The last decade has seen a large increase in the number of detected 
$\gamma$-ray pulsars.  At GeV energies, the number has grown from two
to at least six (and possibly nine) pulsar detections by the EGRET
telescope on the Compton Gamma Ray Observatory (CGRO) (Thompson 2000).  
However, even
with the advance of imaging Cherenkov telescopes in both northern
and southern hemispheres, the number of detections of pulsed emission
at energies above 20 GeV (Weekes et al. 1998) has remained the same (zero), 
or even decreased if one counts the ``detections" of the non-imaging 
telescopes of the eighties.  
In the coming decade, this unexplored region above 20 GeV may
hold the key to a question on which theorists have disagreed for at least
two decades, that of how and where high energy emission emerges from the pulsar
and how it relates to the radio emission.  Furthermore, the known 
$\gamma$-ray pulsars are still a tiny fraction of the known radio pulsars,
of which there are currently over 1000 (Camilo et al. 2000).  The next-generation
$\gamma$-ray telescopes, both in space and on the ground, will not only
be breaching the unexplored territory between 20 and 200 GeV, but are
expected to make an unprecented increase in the $\gamma$-ray pulsar
population.  GLAST alone will probably detect several hundred or so radio-selected
pulsars, with the predicted number being very model dependent.  However,
the number of radio-quiet $\gamma$-ray pulsars could dwarf the number of
radio-selected $\gamma$-ray pulsars and even approach the total radio
pulsar population.

I will give an overview of the current high-energy emission models and
discuss their predictions for emission above 1 GeV.  
Because it is not yet clear how and where in the pulsar magnetosphere 
the non-thermal high-energy radiation originates, two competing models have
developed.  Polar cap models (Daugherty \& Harding 1982, 1996; Usov \& Melrose 1995) 
assume that particles are accelerated above the neutron star surface and that
$\gamma$-rays result from a curvature radiation or inverse Compton induced 
pair cascade in a strong magnetic field.  Outer-gap models (Cheng, Ho \& Ruderman 
1986, Romani 1996, Hirotani \& Shibata 1999) assume that acceleration
occurs along null charge surfaces in the outer magnetosphere and that 
$\gamma$-rays result from photon-photon pair production-induced cascades.
These two types of models and their variations make contrasting predictions
for the numbers of radio-quiet and radio-loud $\gamma$-ray pulsars and of
their spectral characteristics.  

\section*{High-Energy Emission Models}

Since we observe pulsed emission up to 10 GeV in $\gamma$-ray pulsars, there is 
no dispute that particles are accelerated to extremely relativistic energies in
their magnetospheres.  It is also generally agreed that the particle Lorentz factors 
must be in the range of at least $10^5 - 10^7$ and that these energies are the 
result of acceleration by large-scale electric fields.  The source of the field is
no mystery.  Rotating, magnetized neutron stars are natural unipolar inductors, 
generating huge ${\bf v x B}$ electric fields.  However, they are capable of 
pulling charges out of the star 
against the force of gravity (Goldreich \& Julian 1969) and it is believed that the
resulting charge density that builds up in a neutron star magnetosphere
is able to short out the electric field parallel to the magnetic field 
($E_{\parallel}$)(thus allowing the field to corotate with the star) 
everywhere except at
a few locations.  These spots where ${\bf E \cdot B} \neq 0$ are thought to
occur above the surface at the polar caps and along the null charge surface,
${\bf \Omega \cdot B} = 0$, where the corotation charge changes sign.  These are
the purported sites of particle acceleration and have given rise to the 
two classes of high energy emission models.

\begin{figure}[t] 
\vskip -0.7cm
\centerline{\psfig{file=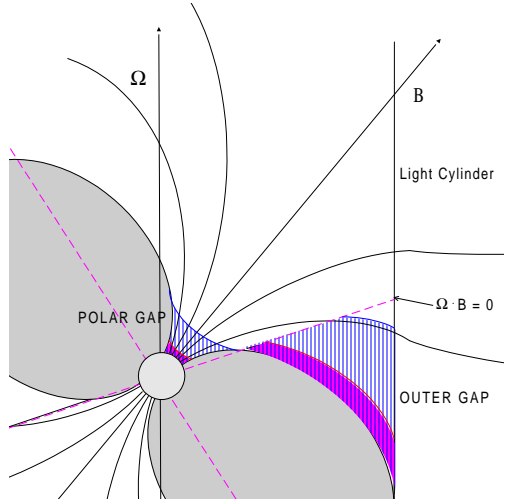,height=9cm}}
\vskip -2.0cm
\caption{Schematic geometry of polar and outer gaps.  Dark solid regions are
thin gaps of younger pulsars.  Hatched regions are thick gaps of older pulsars
(see text).}
\label{fig1}
\end{figure}

\subsection*{Polar cap models}

Polar cap models for pulsar high energy emission date from the early
work of Sturrock (1971) and Ruderman \& Sutherland (1975), who proposed 
particle acceleration and radiation near the neutron star surface at the 
magnetic poles.  There is a large variation among polar cap models, with 
the primary division being whether or not there is free emission of particles 
from the neutron star surface.  This question is still somewhat subject to 
debate, due to our incomplete understanding of the neutron star surface 
composition and physics.  The subclass of polar cap models based on free 
emission of particles of either sign, called space-charge limited flow (SCLF) 
models, assumes that the surface temperature 
of the neutron star (many of which have now been measured in the range 
$T \sim 10^5 - 10^6$ K) exceeds the ion and electron
thermal emission temperatures.  Although $E_{\parallel} = 0$ at the neutron star 
surface in these models, the space charge along open field lines above the surface
falls short of the corotation charge, due to the curvature of the field (Arons 1983)
or to general relativistic inertial frame dragging (Muslimov \& Tsygan 1992).
The $E_{\parallel}$ generated by the charge deficit accelerates particles, which
radiate inverse Compton (IC) photons by resonant scattering of thermal X-rays from the neutron star surface (when they reach energies 
$\gamma \sim 10^2 - 10^6$) and curvature (CR) photons (at energies $\gamma 
\lsim 10^6$).  Both IC and CR photons can produce $e^+ e^-$ pairs in the strong 
magnetic field.  However, it is found (Harding \& Muslimov 2000) that in all
but the very high-field pulsars ($B \gtrsim 10^{13}$ G), 
the IC pair formation fronts do not produce
sufficient pairs to screen the $E_{\parallel}$ or are unstable, due to returning 
positrons which disrupt $E_{\parallel}$ near the surface. 
(Harding \& Muslimov 1998 [HM98]) found in this case that stable acceleration
zones can form at 0.5 - 1 stellar radii above the surface, where the density of
soft X-rays from the neutron star surface decreases and CR photons from
both primary electrons and returning positrons produce stable pair formation 
fronts.  The primary particle energies can then reach $\sim 10^7$ before pair 
production screens the field.  

As the pulsar ages and its period increases, the cascade produces fewer pairs 
and it becomes more difficult to to produce a pair formation front and 
screen the $E_{\parallel}$.  The acceleration zone grows longer and narrower
as the particles must accelerate over larger distance to radiate pair-producing
photons, until pair fronts can no longer form and the pulsar dies as a radio
pulsar.  Thus, as shown in Figure 1, young pulsars will have thin accelerator gaps, 
while old pulsars will
have thick gaps with cascades forming at higher altitudes.

The type of polar cap cascade which produces high-energy radiation depends on
the primary radiation mechanism, which in turn depends on which photons (IC or CR)
control the production of pairs responsible for the screening of the accelerating 
field.  In pulsars where IC-controlled 
acceleration zones are stable, particle energies are limited to Lorentz factors
$10^{5} - 10^{6}$ (HM98) and IC is both the dominant primary radiation mechanism and
the initiator of the pair cascade (Sturner et al. 1995).   
In pulsars where IC photons either
cannot screen the accelerating field or IC-controlled zones are unstable, the
primary particles continue accelerating up to Lorentz factors $\sim 10^7$.
CR is then the dominant primary radiation mechanism and initiates the pair cascade.
In the original version of the CR-initiated polar cap pair cascade (Daugherty 
\& Harding 1982, 1996) the emergent cascade spectrum is dominated by synchrotron
radiation from the pairs and has a very sharp high energy cutoff at several GeV
due to pair production attenuation.  Recently, Zhang \& Harding (2000a) noted that the pairs produced in polar cap
cascades may resonant-scatter the soft thermal photons from the neutron
star surface, losing most of the remaining parallel energy they could not
lose via synchrotron emission. 

\subsection*{Outer gap models}

The outer gap models for $\gamma$-ray pulsars are based on the existence of a vacuum
gap in the outer magnetosphere which may develop between the last open field line and
the null charge surface (${\bf \Omega \cdot B} = 0$) (see Figure 1) in charge 
separated magnetospheres.  
The gaps arise because charges escaping through the light cylinder along open field
lines above the null charge surface cannot be replenished from below.  The first
outer gap $\gamma$-ray pulsar models (Cheng, Ho \& Ruderman 1986 [CHR]) assumed that
emission is seen from gaps associated with both magnetic poles, but this picture, 
although successful in fitting the spectrum of the Crab and Vela pulsars,
did not reproduce the observed pulsar light curves.  More recent 
outer gap models assuming emission from one pole can more successfully reproduce
the observed light curves (Romani \& Yadigaroglu 1995 [RY95]).  Pairs from the polar
cap cascades, which flow out along all the open field lines, will undoubtedly
pollute the outer gaps to some extent, but this effect has yet to be investigated.

The electron-positron pairs needed to provide the current, and therefore allow particle acceleration, in the outer gaps are produced by photon-photon pair production.
In young Crab-like pulsars, the pairs are produced by curvature photons from the primary 
particles interacting with non-thermal synchrotron X-rays from the same pairs.
In older Vela-like pulsars, where non-thermal X-ray emission is much lower, the pairs
were assumed to come from interaction of primary particle inverse Compton photons with 
infra-red photons.  However, this original Vela-type model (CHR) predicted large fluxes 
of TeV emission, from inverse Compton scattering of the infra-red photons by primary 
electrons, which violates the observed upper limits 
(Nel et al. 1993) by several orders of magnitude.  Cheng (1994) revised the outer gap 
model for Vela-type pulsars by proposing another self-sustaining gap mechanism where 
thermal X-rays from the neutron star surface interact with primary radiation to
produce pairs, replacing the infra-red radiation (which has 
also never been observed).  Some of the accelerated pairs flow downward to heat the 
surface and maintain the required thermal X-ray emission.  The modern outer gap Vela-type models (Romani 1996, Zhang \& Cheng 1997) all adopt this picture.

As in polar cap models, it becomes more difficult for older pulsars to produce
the pairs required to screen the field and ``close the gap", so that young
pulsars have thin gaps and old pulsars have thick gaps, as shown in Figure 1.
However, unlike in polar cap (SCLF) models, pair production plays a critical
role in production of the high energy emission: it allows the current to flow and 
particle acccleration to take place in the gap. Beyond a death line in 
period-magnetic field space, and well before the traditional radio-pulsar death line, pairs cannot close the outer gap and the pulsar cannot 
emit high energy radiation.  This outer gap death line for $\gamma$-ray pulsars
falls around $P = 0.3$ s for $B \sim 10^{12}$ G (Chen \& Ruderman 1993), putting 
Geminga just barely among the living.  
The observed non-thermal radiation in Crab-like pulsars is a combination of synchrotron
emission and synchrotron self-Compton emission from pairs.  In Vela-type pulsars, the
non-thermal radiation is a combination of curvature and curvature self-Compton emission 
from the primaries at $\gamma$-ray energies, and synchrotron emission from the pairs at
optical through X-ray energies.  The high-energy spectra in
both types of outer gap model have cutoffs around 10 GeV, due to the 
radiation-reaction cutoff
in the primary particle spectrum, which are much less sharp than the attenuation cutoffs
in polar cap model spectra.  

\section*{Predictions for High Energy Emission}

Observations of pulsars in the unexplored energy region above 20 GeV and more sensitive
measurements above 1 GeV may finally be able to discriminate in favor of polar cap or
outer gap models (or eliminate both!).  I will discuss three areas where future
observations will be able to test distinctive predictions of the models: spectral
high-energy cutoffs, luminosities and population statistics (which radio-selected 
pulsars are $\gamma$-ray loud, which $\gamma$-ray pulsars are radio quiet).  

\subsection*{Spectral shape and cutoffs}

\begin{figure}[t] 
\vskip -1.0cm
\centerline{\psfig{file=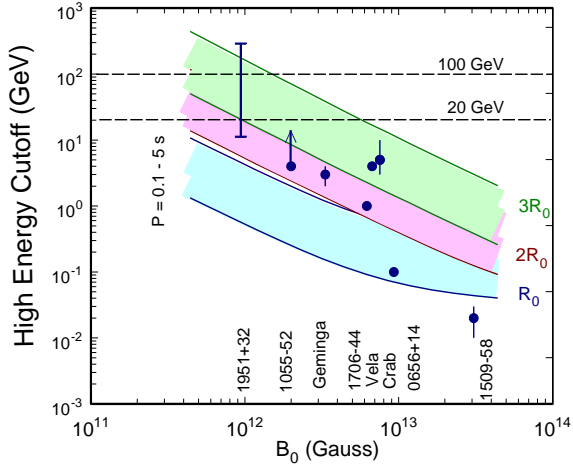,height=8cm}}
\caption{Calculated high-energy spectral cutoff energies due to magnetic 
pair production attenuation vs. surface field strength for a range of periods
at different photon emission radii.  Also shown are measured turnover energies
of detected pulsars.}
\label{fig2}
\end{figure}

Polar cap models predict that the $\gamma$-ray spectra cutoff very sharply (as
a ``super-exponential") due to one-photon pair production attenuation, 
at a field-dependent energy, while outer gap model spectra cut off more slowly
(as a simple exponential) due to a particle acceleration limit.  The highly
relativistic particles emit photons at very small ($\theta \sim 1/\gamma$) angles to
the open magnetic field lines.  The photons of energy $\epsilon$ (in units of 
$mc^2$), emitted near the neutron star surface, are initially below 
the threshold for one-photon pair production ($\epsilon_{th} = 2/\sin\theta$), but
may reach threshold by increasing $\theta$ in the course of propagating across
curved field lines.  The polar cap model $\gamma$-ray spectrum will exhibit 
a cutoff at the 
pair escape energy (cf. Harding et al. 1997 [HBG97]), i.e. the highest energy at which photons
emitted at a given location can escape the magnetosphere without pair producing.
An estimate of this cutoff energy, assuming emission
along the polar cap outer rim, $\theta \simeq (2\pi R / cP)^{1/2}$, at radius $R$, is
(see Zhang \& Harding 2000a, Eqn [28])
\be \label{Ec}
E_c \sim 2\,\,{\rm GeV}\,P^{1/2}\,B_{0,12}^{-1}\,\mbox{\large $\left({R\over R_0}\right)$}^{5/2},  ~~~~~~B_{0,12} \lsim 10\,(R/R_0)^2 \nonumber
\ee
\vskip -0.8cm
\be
E_c \sim 0.2\,\,{\rm GeV}\,P^{1/2}\,\mbox{\large $\left({R\over R_0}\right)$}^{1/2},~~~~~~ B_{0,12} \gsim 10\,(R/R_0)^2 
\ee
where $P$, $R_o$ and $B_{0,12}$ are the neutron star period, radius and surface 
magnetic field in units of $10^{12}$ G.
Figure 2 shows a more accurate calculation of the predicted high-energy cutoff energy 
as a function of surface field strength for different radii of photon emission, 
computed by numerically propagating photons
through a neutron star magnetosphere and taking into account general relativistic effects
of a Schwarzschild metric (as in HBG97).  Also plotted are the
observed cutoff energies of eight $\gamma$-ray pulsars versus their
surface fields derived from $P$ and $\dot P$, assuming $R_0 = 10^6$ cm.  The
very steep spectrum with index 3 measured for PSR0656+14 (Ramanamuthy et al. 1996) is 
assumed to indicate a cutoff around 100 MeV.  
The cutoff energy for the highest field pulsar, PSR1509-58, falls below
the predicted pair escape energy at the surface.  However photon splitting, in which
a single photon splits into two lower energy photons, becomes the
dominant attenuation process in fields above $\sim 2 \times 10^{13}$ G and 
lowers the photon escape energy (HBG97).  There seems to be
an increasing cutoff energy with decreasing surface field in the observed pulsars, with
a dependence even stronger than predicted by the polar cap model for a constant emission
radius.  However, some increase in emission radius is expected due to the trend for
larger acceleration zones in older pulsars, but work to quantify this trend is still
in progress.

\begin{figure}[t] 
\vskip -1.0cm
\centerline{\psfig{file=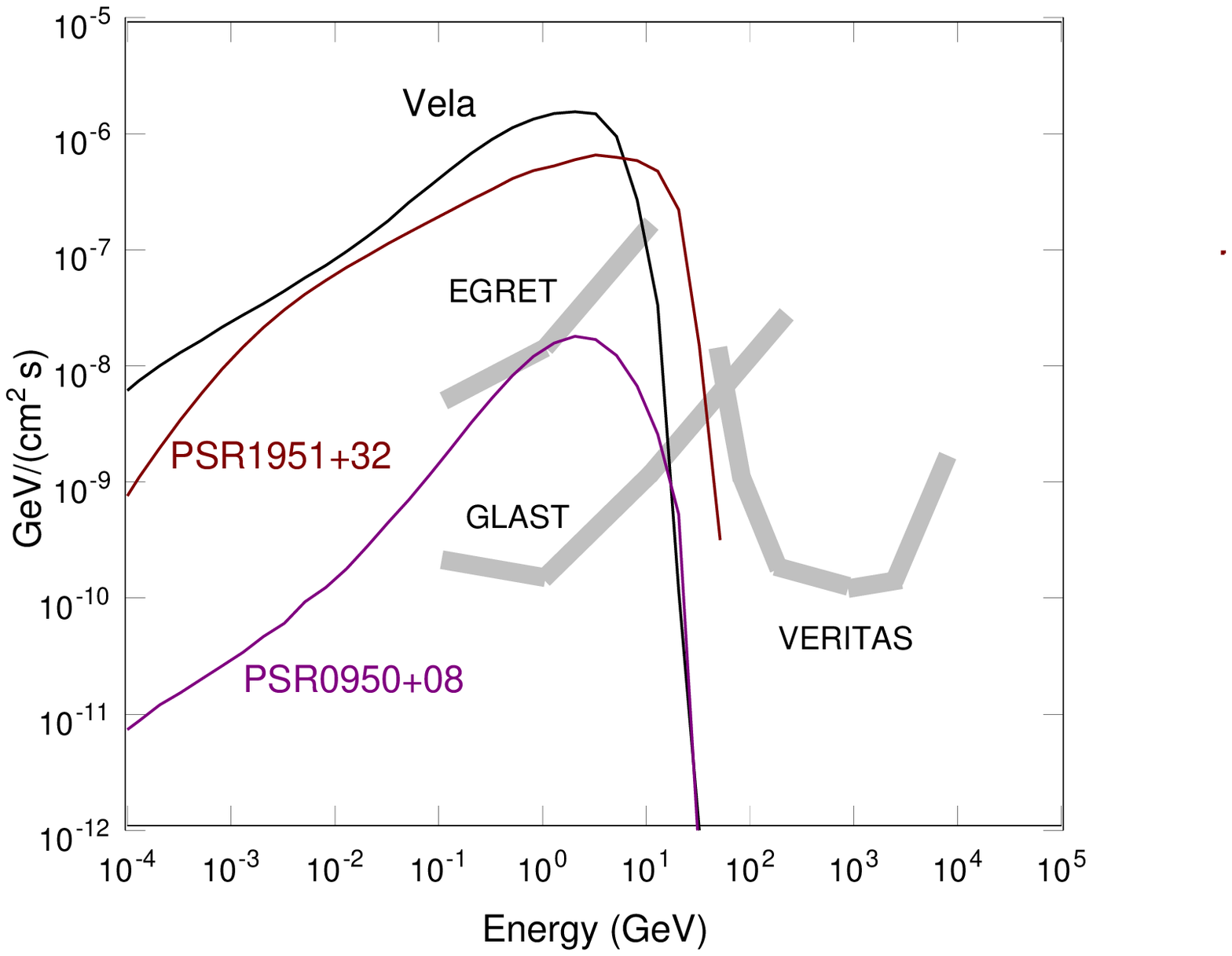,height=8cm}}
\caption{Polar cap model spectra of three different pulsars and sensitivity thresholds
of various detectors. Vela and PSR1951+32
have been detected by EGRET as $\gamma$-ray pulsars.}
\label{fig3}
\end{figure}

From Figure 2, it appears that long period pulsars with low magnetic fields (``old"
pulsars) will be the best candidates for detection above 20 GeV.  However, there 
are two effects which work against detection of these pulsars at $\gamma$-ray energies.
One is that the curvature radiation energy of the primaries decreases with increasing 
period, due to both an increasing radius of curvature of the last open field line
and a decreasing particle acceleration energy.  The maximum CR energy starts to move
below the photon escape energy and thus determines the cutoff energy.
Figure 3 shows cascade model simulations of high-energy spectra for three pulsars
of different types.  The model spectrum of Vela, a young pulsar with a high magnetic 
field and the brightest steady $\gamma$-ray source seen by EGRET, shows the sharp
``super-exponential" ($\exp({-\alpha})$, where the attenuation coefficient $\alpha$ is
itself an exponential of the photon energy) high-energy cutoff below 10 GeV.  
There is some evidence that the observed high-energy cutoffs are indeed steeper than a
simple exponential of the photon energy (Nel \& De Jager 1995).
The spectrum of PSR1951+32,
having a short period but surface field of only $9.8 \times 10^{11}$ G and no detected
high-energy cutoff below 10 GeV, has a predicted sharp cutoff around 20 GeV.  The
spectrum of PSR0950+08, an older pulsar with period $P = 0.253$ s and age $\tau \sim 10^7$
yr, shows
a more gradual high-energy cutoff around 2 GeV, the curvature radiation
critical energy, which steepens at the pair escape energy 
around 20 GeV.  This pulsar was not detected by EGRET, but should be easily detectable
by GLAST.  If the present version of the polar cap model is correct, then pulsed
emission will be difficult to detect with the next generation air-Cherenkov detectors, 
even from short-period, low-field pulsar like PSR1951+32, unless energy thresholds below
50 GeV and preferably 20 GeV can be achieved.  

The picture is quite different in outer gap models (and much more hopeful for
ground-based observers).  When the high-energy photons are emitted in the outer
magnetosphere, where the local magnetic field is orders of magnitiude lower than
the surface field, one-photon pair production plays no role in either the pair cascade or
the spectral attenuation.  In this case the high-energy cutoffs in the photon spectrum 
come from the upper limit of the accelerated particle spectrum, due to radiation reaction.
The shape of the cutoff is thus a simple exponential, more gradual than in polar
cap model spectra.  Figure 4 shows the broad-band outer-gap model spectrum of Vela
(Romani 1996), superposed on the measured spectrum from optical to VHE $\gamma$-rays
and the polar cap model spectrum (Harding \& Daugherty 1996).  The more gradual
high-energy cutoff of the outer gap spectrum relative to that of the polar cap
spectrum is apparent.  However, due to the large errors of the EGRET data points above
1 GeV, the measurements at present to not definitely discriminate between model
spectra.  GLAST should have the energy resolution and dynamic range to measure the shape 
of the cutoffs seen by EGRET and should be able to rule out either the simple exponential 
or super-exponential shape.  In addition, GLAST will detect enough $\gamma$-ray 
pulsars with different field strengths to look for a correlation between surface
field strength and cutoff energy.  

Outer gap models predict an emission component at TeV energies due to inverse Compton 
scattering by gap-accelerated particles.  The original predictions of Cheng et al. (1986) 
were not 
verified by observations of ground-based detectors (Nel et al. 1993), requiring a
revision of the Vela-like model (Cheng 1994).  However, even later models which
predicted lower TeV fluxes (Romani 1996) are above CANGAROO upper limits on pulsed
emission from Vela (see Fig. 4).  The most recent outer gap models (Hirotani 2000), 
have predicted
TeV inverse-Compton fluxes which are below the present observational upper limits,
but which should be detectable with the next generation of TeV detectors.  Unfortunately,
while a TeV emission component is an essential prediction of all outer gap models,
the inverse Compton flux level depends on the pulsed emission spectrum in the 
infra-red (IR) band which is notoriously difficult to measure in most pulsars.  Unmeasured
IR turnovers can decrease the scattered TeV significantly (Romani 2000, priv. comm.).

\begin{figure}[t] 
\vskip -1.0cm
\centerline{\psfig{file=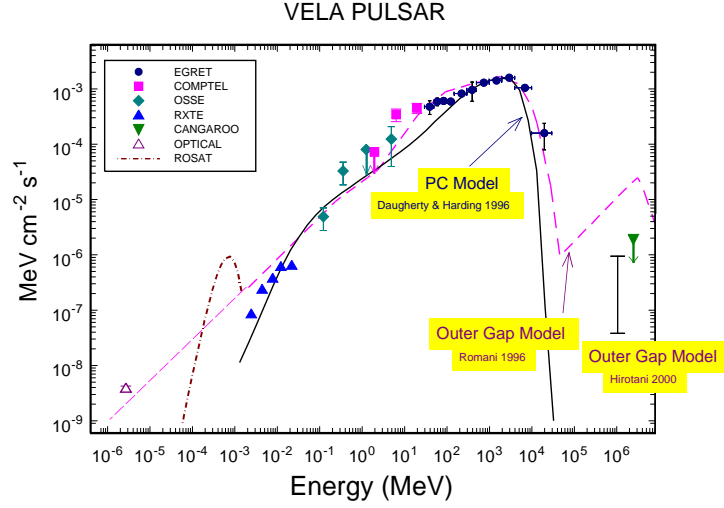,height=8cm}}
\caption{Observed optical to VHE $\gamma$-ray spectrum of the Vela pulsar with
polar cap (solid line) and outer gap (dashed line) model spectra.  Data points 
are from Thompson (2000).}
\label{fig4}
\end{figure}

\subsection*{Luminosities}

Predicted $\gamma$-ray pulsar luminosities and
which radio-selected pulsars will be $\gamma$-ray pulsars will also discriminate
between polar cap and outer gap models.  In polar cap models, the $\gamma$-ray 
luminosity is roughly proportional to the polar cap current of primary particles, 
$N_p \propto B_o P^{-2}$.  The CR-initiated cascade model of Zhang \& Harding (2000a)
predicts that,
\be
L_{\gamma}^{ZH}(I) = 9.4 \times 10^{31}\,B_{12}^{6/7}P^{-13/7}\,{\rm erg\,s^{-1}}
\ee
\vskip -0.5cm
\be
L_{\gamma}^{ZH}(II) = 1.6 \times 10^{31}\,B_{12} P^{-9/4}\,{\rm erg\,s^{-1}} 
\ee
where Regime I applies to young pulsars satisfying
\be
B_{12}^{1/7} P^{-11/28} > 6.0
\ee
and Regime II applies to older pulsars.  The ICS-initiated polar cap cascade model
of Sturner \& Dermer (1994) predicts that
\be
L_{\gamma}^{SD} = 10^{32}\,B_{12}^{3/2}P^{-2}\,{\rm erg\,s^{-1}}
\ee
The predicted $\gamma$-ray luminosity in the outer gap models, on the other hand,  
is not as directly tied to the polar cap current, but rather depends on the fraction
of open field lines (and thus fraction of the polar cap current) that is spanned by
the outer gap accelerator.  The model of Romani \& Yadigaroglu (1995) predicts
\be
L_{\gamma}^{RY} = 2.5 \times 10^{32}\,B_{12}^{0.48}P^{-2.48}\,{\rm erg\,s^{-1}}
\ee
while the outer gap model of Cheng \& Zhang (1998) predicts
\be
L_{\gamma}^{CZ} = 6.3 \times 10^{33}\,B_{12}^{0.3}P^{-0.3}\,{\rm erg\,s^{-1}}.
\ee
The known $\gamma$-ray pulsars, assuming a constant solid angle for all sources,
follow the luminosity dependence, $L_{\gamma} \propto L_{SD}^{1/2} \propto B_0 P^{-2}$,
where $L_{SD}$ is the spin-down luminosity.  The polar cap models thus more
naturally explain this observed dependence. 

Polar cap models predict that all pulsars are capable of $\gamma$-ray emission at
some level.  Which pulsars are detected as $\gamma$-ray pulsars is thus a matter of
sensitivity.  Outer gap models predict a ``death line" for $\gamma$-ray emission in
pulsars, which is a division in period-surface magnetic field space between young
pulsars capable of sustaining pair production (and thus activity) in the outer gaps
from the older pulsars which cannot (Ruderman \& Halpern 1993, Chen \& Ruderman 1993).
Thus, a critical test of outer models is the non-detection of pulsars with ages
much exceeding that of Geminga.

\subsection*{Population statistics and radio-quiet pulsars}

Polar cap and outer gap models predict different ratios of radio-loud to radio-quiet
$\gamma$-ray pulsars, primarily due to the different geometry of the high-energy
emission regions and its location relative to the radio emission region.  Numerous
studies of radio emission morphology of many pulsars (e.g. Rankin 1993, Gil \& Han
1995) argue in favor of an origin in the polar regions, within tens of stellar
radii of the neutron star surface.  Thus, polar cap $\gamma$-ray emission is expected
to have a much higher correlation with radio emission.  In fact, the radio emission
is physically linked to the $\gamma$-ray emission in polar cap models if pairs from
the high-energy cascades are a necessary requirement for coherent radio emission.
On the other hand, the high energy emission in the outer gap is radiated
in a different direction from the radio emission, which allows these models to 
account for the observed phase offsets of the radio and $\gamma$-ray pulses.  At the
same time, there will be fewer radio-$\gamma$-ray coincidences and thus a larger
number of radio-quiet $\gamma$-ray pulsars.  In Romani \& Yadigaroglu's (RY95)
geometrical outer gap model, the radio emission originates from the magnetic pole
opposite to the one connected to the visible outer gap.  Many observer lines-of-sight
miss the radio beam but intersect the outer-gap $\gamma$-ray beam, having a much
larger solid angle.  When the line-of-sight does intersect both, the radio pulse
leads the $\gamma$-ray pulse, as is observed in most $\gamma$-ray pulsars.

Simulations of the radio and $\gamma$-ray pulsar populations in both models
reflect these intuitive ideas.  In a study of outer gap emission based on the model
of RY95, Yadigaroglu \& Romani (1995) find that the number of 
radio-quiet (Geminga-like) pulsars detectable as point sources by EGRET (17) is
much larger than the number of radio-loud $\gamma$-ray pulsars (5).  
Zhang et al. (2000) find a similar ratio of radio-quiet to radio-loud EGRET $\gamma$-ray
pulsars in their outer gap model and also predict that GLAST will detect 80 radio-loud
and 1100 radio-quiet pulsars.  On the other hand, a study of the 
polar cap $\gamma$-ray pulsar population by Sturner \& Dermer (1996) find that 
radio-quiet pulsars constitute only $25\%$ of the $\gamma$-ray pulsars detectable
by EGRET.  Gonthier et al. (2000) have also found a small ratio of radio-quiet to
radio-loud pulsars detectable by EGRET in the polar cap model, $\sim 10\%$.  
However, they have also computed the number of detections expected for GLAST and find
that the situation is reversed, with about 180 radio-loud and 302 radio-quiet pulsars
detectable, at least as point sources (a much smaller number, $\sim 20$, will be detectable
as pulsed sources).  This is because GLAST will be sensitive to 
pulsars at larger distances than the present radio surveys.  All of the 
population studies of polar cap $\gamma$-ray pulsars have assumed that both $\gamma$-ray
and radio emission is beamed with the same direction and solid angle, and studies
including geometry of beams are needed to refine the estimates.

Recently, another possible population of radio-quiet $\gamma$-ray pulsars has been
suggested by Zhang \& Harding (2000b, see also Harding \& Zhang 2000).  According to
the polar cap model (e.g. Daugherty \& Harding 1996), $\gamma$-ray emission occurs
throughout the entire pulse phase.  Primary electrons that initiate pair cascades at
low altitude
continue to radiate curvature emission on open field lines to high altitudes beyond 
the cascade region, producing a lower level of softer off-beam emission.  Due to the 
flaring of the dipole field lines, this emission may be seen over a large solid angle, 
far exceeding that of the main beams.  Since the radio emission is expected to originate
within ten stellar radii of the neutron star surface, it is quite probable to see
off-beam $\gamma$-ray emission and miss the radio beam.  
Zhang \& Harding (2000b) estimate that the probability of detecting such off-beam 
emission is a factor of $\sim 4-5$ times higher than that of the on-beam 
emission.  At least some of the radio-quiet Gould Belt sources detected
by EGRET (Gehrels et al. 2000, Grenier 2000) could therefore be such off-beam gamma-ray 
pulsars.

\section*{Summary}

I have outlined the current predictions of high-energy pulsar emission models  
which can potentially be tested by future instruments having both higher 
sensitivity and larger energy range.  Probably the most discriminating tests will
be measurement of pulsar spectra at energies from 1 GeV to 10 TeV.  In this
range, polar cap models predict steep spectral cutoffs due to magnetic pair
production attenuation and essentially no detectable emission above about 50 GeV
from any pulsar. While GLAST should be able to measure the shape of these cutoffs, 
it is also important for Air Cherenkov detectors to achieve sensitivity at low energies.
Outer gap models predict not only more gradual spectral cutoffs
around 10 GeV, but an inverse Compton component with a peak in power around 1 TeV. 
The predicted flux of this inverse Compton emission is somewhat model dependent,
but should be detectable by future Air Cherenkov detectors.  The presence of such
a component would be very difficult, if not impossible, to explain in polar cap
models. Distributions of radio-loud pulsars detected
as $\gamma$-ray pulsars by GLAST will be able to test predicted $\gamma$-ray luminosity 
dependence on pulsar parameters. In particular, detection of $\gamma$-ray pulsars
older than about $0.5$ Myr will argue strongly for polar cap models. 
The number of radio-quiet pulsars detected by GLAST will be an important diagnostic.  Although both polar cap and outer cap models
expect more radio-quiet $\gamma$-ray pulsars detectable with GLAST,
the outer gap models will always predict larger ratios of radio-quiet to radio-loud
pulsars due to geometry.


\begin{references}

\bibitem{}Arons, J., {\it ApJ}, {\bf 266}, 215 (1983).
\bibitem{} Camilo, F., et al. , in Pulsar Astronomy: 2000 and Beyond -
   IAU Coll. {\bf 177}, 3 (2000).
\bibitem{}Cheng, K. S., {\it Proc. Toward a Major Atmospheric Cherenkov Detector},
  ed. T. Kifune (Tokyo: Universal Academy), 25 (1994).
\bibitem{}Cheng, K. S., Ho, C. \& Ruderman, M. A., {\it ApJ}, {\bf 300}, 500 (1986).
\bibitem{} Chen, K. \& Ruderman, M. A., {\it ApJ}, {\bf 402}, 264 (1993).
\bibitem{}Daugherty, J. K. \& Harding, A. K., {\it ApJ}, {\bf 252}, 337 (1982).
\bibitem{dau94}Daugherty, J.~K., \& Harding A.~K.: {\it ApJ}, {\bf 429}, 325 (1994).
\bibitem{dau96}Daugherty, J.~K., \& Harding A.~K.: {\it ApJ},
   {\bf 458}, 278 (1996).
\bibitem{} Gehrels, N., et al., {\it Nature}, {\bf 404}, 363 (2000).
\bibitem{} Gil, J. A. \& Han, J. L., {\it ApJ}, {\bf 458}, 265 (1996).
\bibitem{} Goldreich, P. \& Julian, W. H. , {\it ApJ}, {\bf 157}, 869 (1969).
\bibitem{} Gonthier, P. G. et al. , in prep (2000).
\bibitem{} Grenier, I. A. et al., {\it A \& A}, in press (2000).
\bibitem{} Harding, A.~K., Baring, M.~G. \& Gonthier, P.~L. , {\it ApJ}, {\bf 476}, 246
   (1997). (HBG97)
\bibitem{} Harding, A. K. \& Muslimov, A. , {\it ApJ}, {\bf 508}, 328 (1998).
\bibitem{} Harding, A. K. \& Muslimov, A. , in prep (2000).
\bibitem{} Harding, A. K. \& Zhang, B. , {\it ApJ Letters}, submitted (2000).
\bibitem{} Hirotani, K. , {\it ApJ}, in press (2000).
\bibitem{} Hirotani, K. \& Shibata, S. , MNRAS, {\bf 308}, 67 (1999).
\bibitem{} Muslimov, A. G. \& Tsygan, A. I., {\it MNRAS}, {\bf 255}, 61 (1992).
\bibitem{} Nel, H. I. et al., {\it ApJ}, {\bf 418}, 836 (1993).
\bibitem{} Nel, H. I. \& De Jager, O. C. , Ap\&SS, {\bf 230}, 299 (1995).
\bibitem{} Ramanamuthy, P. V. et al. , {\it ApJ}, {\bf 458}, 755 (1996).
\bibitem{} Rankin, J. M. , {\it ApJ}, {\bf 405}, 285 (1993).
\bibitem{} Romani, R. W., {\it ApJ}, {\bf 470}, 469 (1996).
\bibitem{} Romani, R. W. \& Yadigaroglu, I.-A., {\it ApJ}, {\bf 438}, 314 (1995).
\bibitem{} Ruderman, M.A. \& Sutherland, P. G., {\it ApJ}, {\bf 196}, 51 (1975).
\bibitem{} Ruderman \& Halpern , {\it ApJ}, {\bf 415}, 286 (1993).
\bibitem{} Sturner, S. J. \& Dermer, C. D.: {\it Ap.\ J.} {\bf 420}, L79 (1994).
\bibitem{} Sturner, S. J. \& Dermer, C. D. , A\&AS, {\bf 120}, 99 (1996).
\bibitem{} Sturrock, P.A. , {\it ApJ}, {\bf 164}, 529 (1971).
\bibitem{} Thompson, D. J., these proceedings (2000).
\bibitem{} Usov, V. V. \& Melrose, D. B., {\it Aust. J. Phys.}, {\bf 48}, 571 (1995).
\bibitem{} Weekes, T. C. et al., in {\it Neutron Stars and Pulsars} ed. N. Shibazaki 
   et al. (Univ. Academy Press, Tokyo), 457 (1998).
\bibitem{} Yadigaroglu, I.-A. \& Romani, R. W., {\it ApJ}, {\bf 449}, 211 (1995).
\bibitem{} Zhang, L. \& Cheng, K. S. , {\it ApJ}, {\bf 487}, 370 (1997).
\bibitem{} Zhang, B. \& Harding, A. K. , {\it ApJ}, {\bf 532}, 1150 (2000a).
\bibitem{} Zhang, B. \& Harding, A. K. , these proceedings (2000b).
\bibitem{} Zhang, L., Zhang, Y. J. \& Cheng, K. S. 2000, A \& A, 357, 957.

\end{references}
\end{document}